\documentclass[conference]{IEEEtran}
\IEEEoverridecommandlockouts

\usepackage{cite}
\usepackage{amsmath,amssymb,amsfonts}
\usepackage{algorithmic}
\usepackage{graphicx}
\usepackage{textcomp}
\usepackage{xcolor}

\usepackage{array}
\usepackage{multirow}
\usepackage{ragged2e}
\usepackage{booktabs}
\usepackage[normalem]{ulem}
\usepackage{tabularray}
\usepackage{tabularx}    
\usepackage{adjustbox}
\usepackage{graphicx}
\usepackage{float}
\usepackage{amssymb}
\makeatletter

\newcommand{\Rmnum}[1]{\expandafter\@slowromancap\romannumeral #1@}
\makeatother

\def\BibTeX{{\rm B\kern-.05em{\sc i\kern-.025em b}\kern-.08em
    T\kern-.1667em\lower.7ex\hbox{E}\kern-.125emX}}
\begin{document}

\title{Qieemo: Speech Is All You Need in the Emotion Recognition in Conversations\\
	
\thanks{* These authors contributed equally to this work.}
}

\author{\IEEEauthorblockN{1\textsuperscript{st} Jinming Chen*}
\IEEEauthorblockA{\textit{Qifu Technology, Inc} \\
Shanghai, China \\
chenjinming-jk@360shuke.com}
\and
\IEEEauthorblockN{2\textsuperscript{nd} Jingyi Fang*}
\IEEEauthorblockA{\textit{Qifu Technology, Inc} \\
Shanghai, China \\
fangjingyi-jk@360shuke.com}
\and
\IEEEauthorblockN{3\textsuperscript{rd} Yuanzhong Zheng}
\IEEEauthorblockA{\textit{Qifu Technology, Inc} \\
Shanghai, China \\
zhengyuanzhong-jk@360shuke.com}
\and
\IEEEauthorblockN{4\textsuperscript{th} Yaoxuan Wang}
\IEEEauthorblockA{\textit{Qifu Technology, Inc} \\
Shanghai, China \\
wangyaoxuan-jk@360shuke.com}
\and
\IEEEauthorblockN{5\textsuperscript{th} Haojun Fei}
\IEEEauthorblockA{\textit{Qifu Technology, Inc} \\
Shanghai, China \\
feihaojun-jk@360shuke.com}
\and
}
\maketitle

\begin{abstract}
Emotion recognition plays a pivotal role in intelligent human-machine interaction systems. Multimodal approaches benefit from the fusion of diverse modalities, thereby improving the recognition accuracy. However, the lack of high-quality multimodal data and the challenge of achieving optimal alignment between different modalities significantly limit the potential for improvement in multimodal approaches. In this paper, the proposed Qieemo framework effectively utilizes the pretrained automatic speech recognition (ASR) model backbone which contains naturally frame aligned textual and emotional features, to achieve precise emotion classification solely based on the audio modality. Furthermore, we design the multimodal fusion (MMF) module and cross-modal attention (CMA) module in order to fuse the phonetic posteriorgram (PPG) and emotional features extracted by the ASR encoder for improving recognition accuracy. The experimental results on the IEMOCAP dataset demonstrate that Qieemo outperforms the benchmark unimodal, multimodal, and self-supervised models with absolute improvements of 3.0\%, 1.2\%, and 1.9\% respectively.
\end{abstract}


\section{Introduction}
Speech emotion recognition (SER) technology is essential in augmenting the interactions naturalness of intelligent human-machine systems by accurate identifying speakers' emotional state \cite{poria2019emotion}. The multimodal emotion recognition methods, which leverage the complementary information provided by different modalities, can achieve higher levels of recognition accuracy compared to unimodal methods \cite{geetha2024multimodal}. The effectiveness of multimodal methods, however, heavily relies on the quality of data acquired from each modality and the alignment of features across them. In conversational scenarios, the text modality typically depends on ASR models which may introduce errors in recognition. The presence of these biases can significantly impair the performance of multimodal methods \cite{Li_Bell_Lai}. Additionally, the alignment among different modalities appears to be crucial in enhancing the accuracy of emotion recognition in multimodal methods. Additionally, these approaches are often constrained to scenarios where they can extract information from speech signals alone, such as voice assistants and telephone customer service. Therefore, our works focus on developing and improving unimodal speech emotion recognition.

In unimodal emotion recognition methods, CNN \cite{ding2024masa,Peng_Lu_Pan_Liu_2021} and RNN \cite{Zhao_Zheng_Zhang_Wang_Zhao_Li_2018} architectures are commonly employed for local feature extraction, while transformer architectures are utilized to capture global information. The CNN-Transformer \cite{tang2024speech} enhances the representation of emotional features in speech by stacking CNN blocks to extract local features and incorporating transformer to extract global features. The MS-SENet incorporates spatial dropout to enhance feature robustness and integrates squeeze-and-excitation (SE) modules to fuse multi-scale features \cite{Li_Wu_Li_Zheng_Wang_Fei_2023}. Besides, some studies utilize various audio features, including mel frequency cepstral coefficents (MFCC), spectrograms and  high-dimensional embedded acoustic information, combined with co-attention mechanisms to improve the emotional representation of acoustic features \cite{Zou_Si_Chen_Rajan_Chng}.

The remarkable performance of speech self-supervised learning in the downstream tasks has paved new avenues for developing SER. B Nasersharif et al. \cite{nasersharif2024exploring} propose a dimension reduction module to apply the output of the wav2vec 2.0 to generate the emotion feature. Qifei Li et al.  \cite{li2024frame} propose a frame-level emotional state alignment method that finetunes the HuBERT model to obtain emotional features. Xinxin Jiao et al. \cite{Jiao_Wang_Yu} propose a novel method called MFHCA, which employs multi-spatial fusion and the HuBERT model to achieve hierarchical cooperative attention on spectrograms and raw audio. The self-supervised learning (SSL) methods leverage extensive speech data to extract sparse vectors, which are subsequently fine-tuned on downstream tasks for the acquisition of emotion-related features. Nevertheless, this approach remains reliant solely on audio modality and lacks the enriched information that can be provided by audio inputs.

\begin{figure*}
\centering
	\includegraphics[scale=.5]{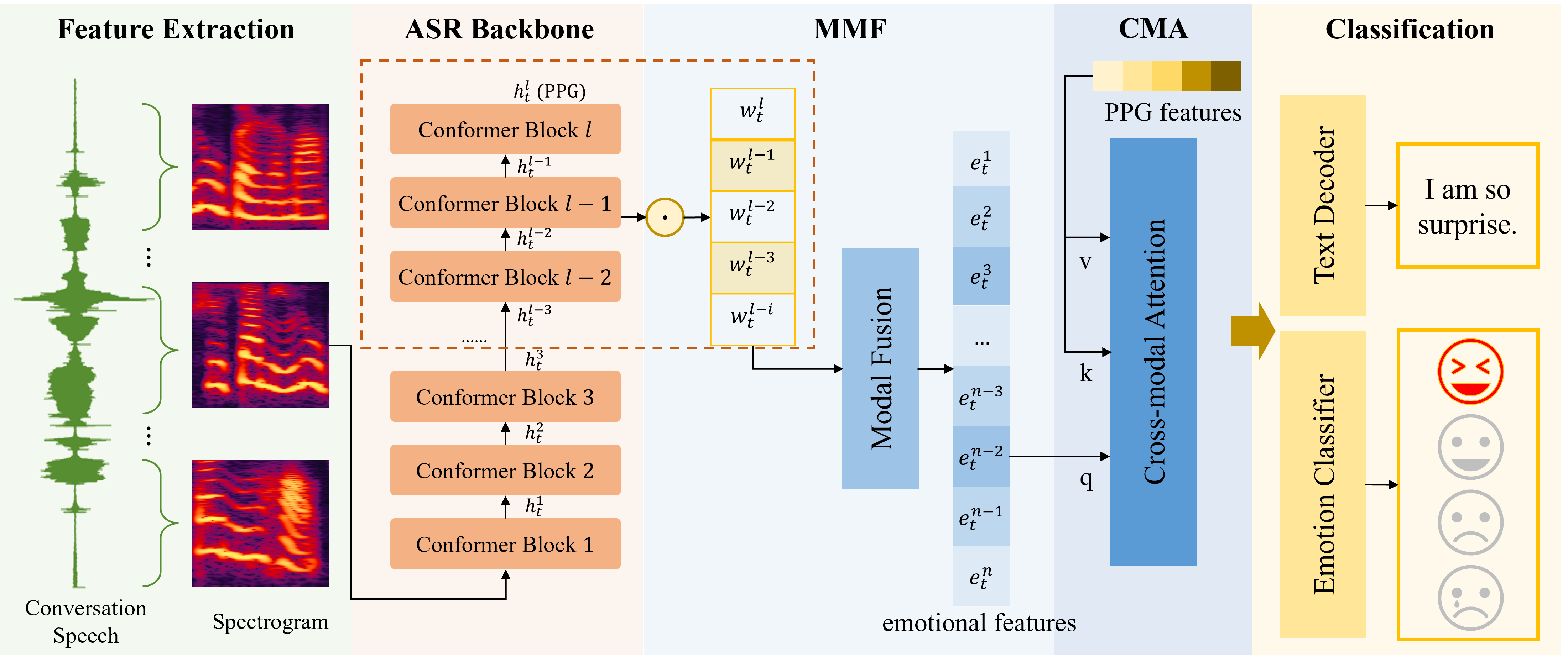}
\caption{Overview of the  proposed Qieemo method}
\label{fig1} 
\end{figure*}

In this paper, we propose a comprehensive framework which can effectively integrates both textual and emotional information extracted from an ASR pretrained model, enabling highly accurate emotion recognition solely based on speech modality input. We propose MMF and CMA modules that combine the text-related PPG features with naturally frame aligned emotional features in order to enhance the robustness of emotion recognition. First, we validated the importance of the features from different layers of the pre-trained ASR encoders for downstream emotion
recognition tasks. It has been proved by experiments that the features from the middle blocks contain strong emotion classification capabilities. Then, by leveraging our proposed MMF and CMA modules, we effectively fuse different blocks of ASR features, resulting in highly expressive emotion representations. Finally, through ablation experiments and contrast experiments on the IEMOCAP dataset, the performance of our unimodal emotion recognition approach has been validated to surpass the state-of-the-art accuracy achieved by multimodal and self-supervised models in emotion recognition. Moreover we prove the universality of Qieemo framework by conducting experiments with different pretrained ASR backbone architectures. This also confirms the feasibility of integrating our emotion recognition scheme into the existing ASR systems. In summary, the main contributions of our work are as follows:
\begin{itemize}
	\item We examine the significant impact of high-dimensional features extracted from different blocks in the pretrained ASR for the task of emotion classification. This discovery lays the foundation for the design of features fusion from pretrained ASR.
	\item We propose the MMF and CMA module to effectively integrate features from different layers of the ASR backbone, demonstrating that this design achieves the highest classification accuracy results on the IEMOCAP dataset compared to multimodal and self-supervised models.
	\item Our proposed Qieemo method is proved to be a universal framework capable of achieving high-accuracy emotion recognition with different ASR backbones, which highlighting its potential integration ability in current ASR systems for emotion recognition in conversation (ERC).
\end{itemize}

\section{Proposed method}
As shown in Fig. \ref{fig1}, we provide an overview of the proposed Qieemo framework for ERC. Based on the ASR Backbone, the design of MMF and CMA module enables the model to obtain accurate emotional labels relying only on the input audio. In Section 2.A, we describe the pretrained ASR backbone in detail. Then MMF and CMA module is introduced in Section 2.B and 2.C. The Multi-stage training strategy is presented in Section 2.D.

\subsection{Pretrained ASR Backbone}
The pretrained ASR structure utilizes an end-to-end attention-based encoder-decoder (AED) framework to extract multi-dimensional speech features from the spectrograms. Specifically, we employ the efficient conformer model as ASR backbone for extracting PPG features. The diagram in Fig. \ref{fig1} illustrates that the encoder of the efficient conformer model is comprised of multiple conformer blocks \cite{Li_Xu_Zhang_2021}. Each conformer block consists of a multi-head self-attention module (MHSA), a convolution module and two feed-forward (FFN) modules. The multi-head self-attention module effectively captures global information, while the convolution module has a natural advantage in extracting local information. More significantly, the efficient conformer introduces progressive down-sampling operation to the conformer encoder, which effectively reducing temporal redundancy caused by the speech frame step and enhancing the extraction of broader acoustic features.

Regarding the spectrogram feature $S_{t}$ of the input audio $x_{t}$, the frame-level acoustic features $H_{t}=\left \{ h_{t}^{1},h_{t}^{2},…,h_{t}^{l}\right \}$ extracted by a series of conformer blocks can be computed as follows:
\begin{equation}
	h_{t}^{\mathrm{FFN}_{1}} = S_t+\frac{\mathrm{FFN_{1} } (S_t)}{2} 
\end{equation}
\begin{equation}
	h_{t}^{\mathrm{MHSA}} =h_{t}^{\mathrm{FFN} _{1}} +{\mathrm{MHSA} (h_{t}^{\mathrm{FFN} _{1}})}
\end{equation}
\begin{equation}
	h_{t}^{\mathrm{Conv} } = h_{t}^{\mathrm{MHSA}} + {\mathrm{Conv} (h_{t}^{\mathrm{MHSA}})}
\end{equation}
\begin{equation}
	h_{t}^{\mathrm{FFN} _{2}} = h_{t}^{\mathrm{Conv} } + \frac{{\mathrm{FFN_{2} } (h_{t}^{\mathrm{Conv} })}}{2} 
\end{equation}
\begin{equation}
	H_{t} = {{\mathrm{LayerNorm} (h_{t}^{\mathrm{FFN} _{2}})}}
\end{equation}
where $l$ denotes the quantity of conformer blocks employed in the efficient conformer model.
\subsection{Multimodal Fusion Module}
The multi-dimensional acoustic features $H_{t}$ extracted from different conformer blocks are believed to possess varying levels of input audio representation, including textual information and a diverse range of linguistic cues, encompassing emotional information. The output feature of the final conformer block $h_{t}^{l}$, which contains crucial text-related information and is commonly named as PPG features, is fed into the ASR decoder. 

To achieve optimal accuracy in emotion classification task, we design a MMF module that effectively combines PPG features and emotional features from different conformer blocks extracted by the ASR Encoder. As shown in Fig. \ref{fig1}, a weighted concatenation method is used as the input for the subsequent  modal fusion. The integration of features from different weighted blocks through the application of Conv2D enables the extraction of utterance-level emotional features $E_{t}=\left \{ e_{t}^{1},e_{t}^{2},…,e_{t}^{n}\right \}$, thereby facilitating the classification of various emotions in conversation corpus.

\subsection{Cross-modal Attention Module}
The text mode is considered to play a crucial role in the recognition of emotions during conversations among various multimodal solutions. Due to the utilization of the pretrained ASR backbone, we are able to extract PPG textual features from the output of the ASR encoder. This allows us to effectively employ these features without relying on any text modal input, thereby achieving emotion recognition solely through unimodal input.

In order to leverage the textual information embedded in the PPG features and enhance the representation of emotional features, we propose a CMA module to enhance the integration of emotional features with PPG features generated by the ASR encoder. Specifically, the emotional features $E_{t}$ are utilized as queries while the PPG features $h_{t}^{l}$ function as keys and values. Both sets of features are fed into the CMA to compute their feature correlations as follows:
\begin{equation}
	Q_{t}  =W_{t}^{Q} E_{t}, K_{t}=W_{t}^{K}h_{t}^{l}, V_{t}=W_{t}^{V}h_{t}^{l}
\end{equation}
\begin{equation}
	Q_{t}^{Att} = \mathrm{Relu} (\mathrm{Softmax} (\frac{Q_{t}(K_{t})_{}^{T}}{\sqrt{d_{att}} } V_{t})
\end{equation}
\begin{equation}
	O_{t}^{Att} = \mathrm{Relu} (\mathrm{Softmax} (\frac{Q_{t}^{Att}(K_{t})_{}^{T}}{\sqrt{d_{att}} } V_{t})
\end{equation}
where $ W_{t}^{Q} $, $ W_{t}^{K} $ and $ W_{t}^{V} $ are trainable weight matrix, the division of the similarity matrix and $  \sqrt{d_att} $ in (7) and (8) contributes to steady gradient descent while training.

\subsection{The Multi-stage Training Strategy}
In the first training stage, we train the ASR backbone model with a substantial amount of speech-text paired data, while freezing the MMF and CMA modules. The pretrained ASR model is employed as the initial model in the second training stage, and it is jointly trained with the MMF and CMA modules using emotion-labeled speech data. By utilizing the ASR pretrained model, which has been trained on extensive speech-text paired data, we can progressively enhance the generation ability of emotion features and consequently obtain a highly accurate emotion classification model.

\section{Experiment}
\subsection{Datasets and Training Method}
IEMOCAP \cite{Busso_Bulut_Lee_Kazemzadeh_Mower_Kim_Chang_Lee_Narayanan_2008} is a well-known multimodal emotion corpus, including audio, visual and text modalities. The experiments in our work exclusively employ audio data as the input modality. The audio corpus comprises five separate recording sessions, each of which showcases a male and a female speaker. To mitigate the potential disclosure of speaker identities and labels, we implement a five-fold cross-validation methodology, wherein one session was excluded in each iteration. Our experiments conduct involved 5531 audio utterances, which are categorized into four emotional states: happy (including both happiness and excitement), sad, neutral, and angry \cite{Guo_Wang_Xu_Dang_Chng_Li_2021}. As for the performance evaluation of SER, we apply three commonly used  metrics \cite{ma2023emotion2vec}: weighted accuracy (WA), unweighted accuracy (UA) and weighted average F1 (WF1).

To demonstrate that Qieemo achieves superior results among unimodal, multimodal, and unsupervised approaches, we select the following benchmark models representing each category. MS-SENet \cite{li2024ms} was chosen for unimodal speech-based emotion classification, MSMSER \cite{Wang_Ma_Ding} for multimodal, and Emotion2Vec \cite{ma2023emotion2vec} for self-supervised emotion classification models. Each model has achieved high recognition accuracy within its respective framework.

\subsection{Results and Comparison}
Firstly we pretrain the ASR backbone model without MMF and CMA modules on the LibriSpeech English dataset. After completing the first stage training, the features extracted from the ASR encoder are utilized as input to the downstream emotion classifier, which is constructed with 256 dimensional linear layers. As illustrated in Table \Rmnum{1}, among block 6 to 12, layer 9 achieves the highest emotion classification performance, with WA, UA and WF1 scores of 67.80 69.72 and 67.52 respectively. The analysis reveals that the accuracy of emotion recognition increases with the depth of ASR encoder initially, reaching its peak at block 9, and then declines. By contrast, the PPG features from the last ASR block contain more textual information, which results in lower accuracy for the emotion classification. Therefore, we can conclude that the highest emotional representation can be attributed to the features extracted from block 9, while the PPG features perform better in representing textual information. We also utilize features extracted from self-supervised models pretrained on the same dataset to implement downstream emotion classification. Our pretrained ASR encoder exhibits a recognition accuracy that is only 0.78\% lower than that of data2vec 2.0. Furthermore, emotion2vec, which extra pretrained on an emotion-labeled corpus, achieves the highest recognition accuracy in our experiments. 

Secondly, the pretrained ASR model is finetuned together with the MMF and CMA modules on the IEMOCAP dataset, where the MMF module integrates features from the last six ASR encoder blocks, while the CMA module achieves frame-level fusion of textual and emotional features. As shown in Table \Rmnum{2}, our approach achieves a WA of 76.42, an UA of 77.71 and a WF1 of 76.20. The performance exceeds the unimodal speech input method MS-SENet (73.38\% WA) by 3.04\% (absolute) and surpasses the multimodal input method MSMSER (75.2\% WA) by 1.22\% (absolute). Notably, Qieemo finetuned results on the IEMOCAP dataset, using only speech modality input, still exceed the current state-of-the-art SSL emotion recognition methods on IEMOCAP downstream tasks (74.48\% WA) by 1.94\% (absolute). The results demonstrate that Qieemo achieves the highest level of recognition accuracy among unimodal, multimodal and self-supervised model downstream fine-tuning scenarios.

\subsection{Ablation Study}
In order to verify the rationality of Qieemo framework design, we conduct the following ablation experiments. Firstly, we validate the proposed MMF and CMA modules in extracting features for enhancing the accuracy of emotion recognition. When removing the CMA module, the WA, UA and WF1 respectively drop to 74.71\%, 75.33\% and 74.74\%, which demonstrates that the frame-level cross-modal fusion between PPG features and emotion features enhances emotion classification accuracy. Additionally, when we used only PPG features from the final block without incorporating the MMF module, the WA, UA and WF1 results drop to 74.29\%, 75.16\% and 74.20\%. Besides, the UA absolutely decreases by 3.42\% when the features from different blocks of the ASR pretrained encoder are not fused.

Secondly, we demonstrate the pretraining process of the ASR backbone also facilitates the extraction of emotional feature representation and implicit textual information. The WA, UA and WF1 results of training the same Qieemo framework on IEMOCAP without utilizing the weights of a pretrained ASR encoder indicate a significant decrease to 57.57\%, 59.60\%, and 56.88\% respectively, which are considerably lower compared to the results achieved using emotional features and textual features from the pretrained ASR encoder.

Finally, in order to demonstrate the universality of our proposed Qieemo framework, we also employed an ASR encoder scheme without progressive down-sampling structure between conformer blocks \cite{Gulati_Qin_Chiu_Parmar_Zhang_Yu_Han_Wang_Zhang_Wu_et} as a pretrained ASR backbone 2. The WA, UA and WF1 results presented in Table \Rmnum{2} indicate that the variations in pretrained backbones have only a 0.44\% difference on the WA of emotion recognition, surpassing all other compared approaches, thereby demonstrating the method’s generalizability.

\begin{table}
	\centering
	\caption{SER task performance of different SSL pretrained models on the IEMOCAP dataset.}
	\renewcommand{\arraystretch}{1.6}
	\resizebox{\linewidth}{!}{
		\begin{tabular}{|c|c|c|c|c|} 
			\hline
			\begin{tabular}[c]{@{}c@{}}\textbf{Self-supervised}\\\textbf{Model}\end{tabular} & \begin{tabular}[c]{@{}c@{}}\textbf{Pre-training }\\\textbf{Corpus}\end{tabular} & \begin{tabular}[c]{@{}c@{}}\textbf{Upstream }\\\textbf{Params}\end{tabular} & \begin{tabular}[c]{@{}c@{}}\textbf{Down}\\\textbf{stream}\end{tabular} & \begin{tabular}[c]{@{}c@{}}\textbf{WA}\\\textbf{(\%)}\end{tabular}  \\ 
			\hline
			wav2vec 2.0 \cite{baevski2020wav2vec20frameworkselfsupervised}                                                                     & \multirow{4}{*}{LS-960}                                                         & 95.04M                                                                      & \multirow{12}{*}{linear}                                               & 63.43                                                               \\ 
			\cline{1-1}\cline{3-3}\cline{5-5}
			HuBERT \cite{Hsu_Bolte_Tsai_Lakhotia_Salakhutdinov_Mohamed_2021}                                                                          &                                                                                 & 94.68M                                                                      &                                                                        & 64.92                                                               \\ 
			\cline{1-1}\cline{3-3}\cline{5-5}
			WavLM \cite{chen2022wavlm}                                                                           &                                                                                 & 94.70M                                                                      &                                                                        & 65.94                                                               \\ 
			\cline{1-1}\cline{3-3}\cline{5-5}
			data2vec 2.0 \cite{baevski2023efficient}                                                                    &                                                                                 & 93.78M                                                                      &                                                                        & \uline{68.58}                                                       \\ 
			\cline{1-3}\cline{5-5}
			emotion2vec \cite{ma2023emotion2vec}                                                                     & LS-960+Emo-262                                                                  & 93.79M                                                                      &                                                                        & \textbf{71.79}                                                      \\ 
			\cline{1-3}\cline{5-5}
			ASR Encoder-\{block 6\}                                                          & \multirow{7}{*}{LS-960}                                                         & \multirow{7}{*}{49.35M}                                                     &                                                                        & 66.64                                                               \\ 
			\cline{1-1}\cline{5-5}
			ASR Encoder-\{block 7\}                                                          &                                                                                 &                                                                             &                                                                        & 67.71                                                               \\ 
			\cline{1-1}\cline{5-5}
			ASR Encoder-\{block 8\}                                                          &                                                                                 &                                                                             &                                                                        & 66.96                                                               \\ 
			\cline{1-1}\cline{5-5}
			ASR Encoder-\{block 9\}                                                          &                                                                                 &                                                                             &                                                                        & \textbf{67.80}                                                      \\ 
			\cline{1-1}\cline{5-5}
			ASR Encoder-\{block 10\}                                                         &                                                                                 &                                                                             &                                                                        & 67.69                                                               \\ 
			\cline{1-1}\cline{5-5}
			ASR Encoder-\{block 11\}                                                         &                                                                                 &                                                                             &                                                                        & 65.31                                                               \\ 
			\cline{1-1}\cline{5-5}
			ASR Encoder-\{layer 12\}                                                         &                                                                                 &                                                                             &                                                                        & 65.66                                                               \\ 
			\hline
			\multicolumn{5}{l}{\begin{tabular}[c]{@{}l@{}}LS-960 indicates LibriSpeech 960 hours, Emo-262 indicates 262 hours emotion data. \\ASR encoder-\{layer *\} indicates features extracted from different blocks of ASR.\end{tabular}}                                                             
		\end{tabular}
	}
	\vspace{-3em}
\end{table}

\begin{table}[H]
	\centering
	\vspace{-1.0em}
	\caption{Ablation experiments of Qieemo on the IEMOCAP dataset, comparing different benchmark models across modalities. }
	\renewcommand{\arraystretch}{1.6}
	\resizebox{\linewidth}{!}{
	\begin{tabular}{|c|c|c|c|c|} 
		\hline
		\textbf{Modality} & \textbf{Method}           & \textbf{UA (\%)} & \textbf{WA (\%)} & \textbf{WF1 (\%)}                                                                                                                                                                         \\ 
		\hline
		S                 & MS-SENet \cite{li2024ms}                 & 73.67            & 73.38            & -                                                                                                                                                                                         \\ 
		\hline
		S                 & SenseVoice-L \cite{speechteam2024funaudiollm}             & 73.9             & 75.3             & 73.2                                                                                                                                                                                      \\ 
		\hline
		S                 & emotion2vec\_large \cite{ma2023emotion2vec}       & 70.7             & 67.3             & 68.5                                                                                                                                                                                      \\ 
		\hline
		S                 & emotion2vec+fintune \cite{ma2023emotion2vec}      & -                & 74.48            & -                                                                                                                                                                                         \\ 
		\hline
		S                 & MFHCA \cite{Jiao_Wang_Yu}                    & 74.57            & 74.24            & -                                                                                                                                                                                         \\ 
		\hline
		S+T               & MSMSER \cite{Wang_Ma_Ding}                   & 76.4             & 75.2             & -                                                                                                                                                                                         \\ 
		\hline
		V+S               & MultiMAE-DER-FSLF \cite{Xiang_Lin_Wu_Bai_2024}        & 63.21            & 63.73            & -                                                                                                                                                                                         \\ 
		\hline
		V+S+T             & TelME \cite{yun2024telme}                    & -                & -                & 70.48                                                                                                                                                                                     \\ 
		\hline
		S                 & Qieemo-\{ASR backbone 2\} & \uline{76.79}    & \uline{75.98}    & \uline{75.89}                                                                                                                                                                             \\ 
		\hline
		S                 & Qieemo-\{ASR backbone 1\} & \textbf{77.71}   & \textbf{76.42}   & \textbf{76.20}                                                                                                                                                                            \\ 
		\hline
		\multicolumn{5}{l}{\begin{tabular}[c]{@{}l@{}}S indicates speech modality, V indicates visual modality, T indicates text modality.\\ "ASR-*" indicates different ASR pretrained backbones.\end{tabular}}   
	\end{tabular}
	}
\end{table}

\section{conclusion}
In this paper, we propose the Qieemo framework based on a pretrained ASR backbone which effectively utilizes the emotion and PPG features from different blocks of ASR encoder. The integration design of MMF and CMA modules facilitates the model in achieving an effective fusion of naturally-aligned emotional and textual features, thereby mitigating the challenges associated with modality alignment in multimodal systems and addressing biases introduced by ASR. Through multiple experiments on IEMOCAP dataset, Qieemo attains the highest recognition accuracy among unimodal, multimodal and self-supervised models. This framework also demonstrates its substantial practicality in emotion recognition in real-time conversation.

\bibliographystyle{IEEEtran}

\bibliography{IEEEabrv,IEEEexample}

\end{document}